\documentclass[%
prl,%
amsmath,amssymb,%
twocolumn%
]{revtex4-2}

\usepackage{graphicx}
\usepackage{dcolumn}
\usepackage{bm}

\usepackage[utf8]{inputenc}
\usepackage[T1]{fontenc}
\usepackage{newtxtext}
\usepackage[varvw]{newtxmath}

\usepackage{color}

\usepackage{bm}
\usepackage{amsmath}
\usepackage{amsfonts}
\usepackage{amssymb}
\usepackage{bm}
\usepackage{dsfont}

\renewcommand{\vec}[1]{\ensuremath{\bm{#1}}}

\newcommand*{\noaddvspace}{\renewcommand*{\addvspace}[1]{}}
\addtocontents{lof}{\protect\noaddvspace}

\newcommand{\ko}{{k_\perp}}
\newcommand{\kz}{{k_z}}
\newcommand{\bv}{{\bm v}}
\newcommand{\bx}{{\bm x}}

\begin{document}

\title{High frequency non-gyrokinetic turbulence at tokamak edge parameters}

\author{M. Raeth}
\email{Mario.Raeth@ipp.mpg.de}
\affiliation{Max Planck Institute for Plasma Physics, Boltzmannstr. 2, 85748 Garching, Germany}

\author{K. Hallatschek}
\email{Klaus.Hallatschek@ipp.mpg.de}
\affiliation{Max Planck Institute for Plasma Physics, Boltzmannstr. 2, 85748 Garching, Germany}

\date{\today}

\begin{abstract}
  First of a kind 6D-Vlasov computer simulations of high frequency ion Bernstein
  wave turbulence for parameters relevant to the tokamak edge show transport
  comparable to sub-Larmor-frequency gyrokinetic turbulence. The customary
  restriction of magnetized plasma turbulence studies to the gyrokinetic
  approximation may not be based on physics but only a practical constraint due
  to computational cost.\end{abstract}
\maketitle

Five-dimensional gyrokinetic turbulence simulations have become the gold
standard for theoretical transport studies in the core of magnetic confinement
systems.  Yet, their descriptive power still falls short of the tokamak
edge-plasma and particularly the high confinement mode (H-mode) transition,
likely due to the presence of high gradients ($\rho_i/L_n=O(1)$) and elevated
fluctuation levels ($\delta n/n=O(1)$). On the other hand, the axioms of
gyrokinetics, that the distribution function is nearly constant on a Larmor
circle and the turbulence time scales long compared to the Larmor period, are
ultimately just assumptions.
Time and again, confined plasma has proven to evade preconceptions, such as
with the unexpected finding of the importance of ``hyperfine turbulence''
\cite{jenko2004nature} on the extremely small electron Larmor radius scale.
Nevertheless, owing to the many extremely disparate time scales for strongly
magnetized plasmas (e.g. for transport, gyration, Alfven-waves), there
have been only steps in the direction of full 6D simulations, consisting of
only two space dimensions \cite{deng2016cyclokinetic} and/or linear
verification with standard gyro-kinetic ion temperature gradient (ITG) modes
\cite{sturdevant16}, or the study of waves \cite{yu2022verification} in a
gyrokinetic background. (The situation is very different from astrophysics or
inertial fusion, where 6D simulations are common place.)

Using the novel 6D turbulence code BSL6D, which is particularly efficient in
simulating ion gyrations \cite{kkormann2019}, we demonstrate that indeed
genuinely non-gyrokinetic instabilities and turbulence have to be taken into
account for high gradients such as those typical for a tokamak edge.


The target of our 6D simulations is the ion kinetic equation in dimensionless
($\rho_i = v_{\text{th}}=n=T=1$) variables
\begin{equation}
  \partial_t f+\vec v\cdot\nabla f+(-\nabla\phi+\vec v\times\vec  B)
  \cdot\nabla_{\vec v} f=0\label{eq:vlasov}
\end{equation}
where for simplicity we restrict ourselves to electrostatic perturbations in
``slab geometry'', i.e., to a homogeneous constant magnetic field
$\vec B=\hat z$. Furthermore, the system is assumed to be quasineutral
($n_i=n_e$) and the electrons to be adiabatic, $n_e=\phi$.
The initial state is chosen to be in (unstable) equilibrium, which is why it
has to be a function of the gyrocenter coordinate
$\vec r_{gc}=\vec r-\vec \rho$, $\vec \rho=\hat z\times\vec v$, in particular,
$x_{gc}=x+v_y$. We choose the gradients to be in x direction, such as
\begin{equation}
  f_0:=f|_{t=0}=g(x_{gc},v),\quad g(x,v)=f_M(n(x),T(x)), \label{eq:init}
\end{equation}
with the Boltzmann distribution
$f_M(n,T)=n (2\pi T)^{-3/2}e^{-\frac{v^2}{2T}}$, where $n(x),T(x)$ represent
the background density and temperature profiles as function of the gyrocenter
position.

\paragraph*{Linear equations} To study the stability of gyrokinetic and
non-gyrokinetic modes for given density and temperature gradients we linearize
(\ref{eq:vlasov}) around $h=0$ with $f=h-\phi g_0+f_0$, $g_0=f_M(1,1)$,
resulting in
\begin{equation}
  \partial_t h+\vec v\cdot\nabla h+\vec v\times\hat z\cdot\nabla_{\vec v}
  h=\partial_t \phi g_0+\nabla\phi\cdot\nabla_v (f_0-g_0).\label{eq:linearvlasov}
\end{equation}
Using the Boussinesq approximation
\begin{equation}
  \nabla_v (f_0-g_0)\approx\nabla_v (f_0-g_0)|_{x_{gc}=0}=:\vec v^*g_0,
  \label{eq:boussinesq}
\end{equation}
\begin{equation}
  \text{with} \; \vec v^*=\hat y\left[ \kappa_T\partial_T(\ln f_M)+ \kappa_n\right],
\end{equation}
where $\kappa_n=L_n^{-1}=(\ln n)'$, $\kappa_T=L_T^{-1}=(\ln T)'$,
$\partial_T\ln f_M=(v^2-3)/2$, the linearized equation becomes translation
invariant and the response of the velocity integral over $h$ to an electric
potential on the right hand side of (\ref{eq:linearvlasov}) can be analyzed in
Fourier space.

At $\vec v^*=0$ the response is the well-known Gordeyev integral
\cite{Gordeev1952,brambilla1998kinetic}
\begin{equation}
  \int \frac{h}{\phi}d^3v=G_\omega(k_\perp\sqrt{T},k_z\sqrt{T})=
\end{equation}
\begin{equation}
 = -i\omega \int_0^\infty e^{-2Tk_\perp^2\sin^2\frac{t'}2-\frac{k_z^2Tt'^2}2+i\omega t'} dt',
\end{equation}
where the $T$ dependence has been explicitly indicated. Noting the structure
of the right hand side, the response for finite gradient term $\vec v^*$ can
be expressed using a $T$-derivative,
\begin{equation}
  \int\frac{h}{\phi}d^3v=\left[1-\frac{k_y}{\omega}(\kappa_T\partial_T+\kappa_n)\right]G_\omega(k_\perp\sqrt{T},k_z\sqrt{T})=
\end{equation}
\begin{equation}
  =\left[-\omega+{k_y}(\kappa_T\partial_T+\kappa_n)\right]
  \sum_{p\in\mathds{Z}}\frac{\Gamma_p(T\ko^2)}{|\kz|\sqrt{2T}}Z\left(\frac{\omega-p}{|\kz|\sqrt{2T}}\right),
  \label{eq:ionresponse}
\end{equation}
where the series expansion \cite{brambilla1998kinetic} of $G_\omega$ has been
inserted, $\Gamma_p(x)=I_p(x)e^{-x}$,
$Z(x)=e^{-x^2}(i\sqrt\pi-2\int_0^x e^{x'^2}dx')$ is the Fried-Conte plasma
dispersion function, and the expressions are to be evaluated at $T=1$.
The electrostatic dispersion relation follows from quasi-neutrality
$n_i=n_e$, where $n_i=\int h d^3v-\phi$ and $n_e=\phi$, as
\begin{equation}
  2=\int
    hd^3v\,/\,\phi.\label{eq:dispersionrelation}
\end{equation}

\begin{figure}
  \includegraphics*[width = 0.4 \textwidth]{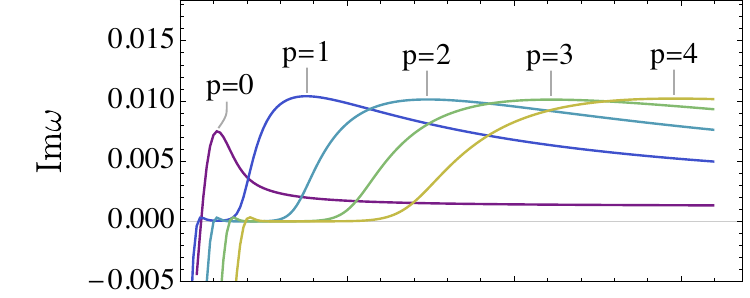}
  \includegraphics*[width = 0.4 \textwidth]{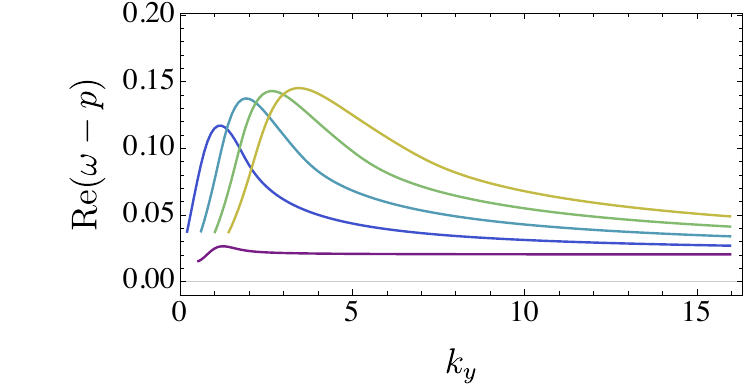}
  \caption{Growth rate $\Im\omega$ and frequency shift $\Re(\omega-p)$
    relative to the associated harmonic $p$ of the Larmor frequency for
    various eigenmodes of (\ref{eq:dispersionrelation}) for $\kappa_T=0.34$,
    $\kappa_n=0.23$, $\ko=k_y$, $k_z= 0.025$; $p=0$ represents the gyrokinetic
    slab ITG mode, $p>0$ non-gyrokinetic unstable
    IBWs. \label{fig_dispersion_relation}}
\end{figure}

In the absence of gradients this equation describes non-gyrokinetic
``neutralized'' ion Bernstein waves (IBW) close to the harmonics of the Larmor
frequency with a frequency shift depending on the wavenumbers
\cite{schmitt1973dispersion}. Already for relatively small gradient parameters
$\kappa_n,\kappa_T<1$ IBW for many harmonics are destabilized, as depicted by
the numerical solutions of (\ref{eq:dispersionrelation}) in
Fig.~\ref{fig_dispersion_relation}, easily exceeding the growth rate of the
(gyrokinetic) ITG mode. The IBWs at higher harmonics tend to have a higher
wavenumber of maximum growth rate.

\paragraph*{Critical gradients of ion Bernstein waves}
In principle, unstable non-gyrokinetic modes can exist for arbitrary complex
$\omega$, since the real and imaginary part of (\ref{eq:dispersionrelation})
can always be satisfied by choosing suitable real parameters
$\kappa_n,\kappa_T$, which in general turn out to be of order of the inverse
Larmor radius. However, we would like to know under which conditions there are
also low gradient non-gyrokinetic instabilities with
$|k_y\kappa_n|,|k_y\kappa_T|\ll1$ and $\omega$ arbitrarily close to the Larmor
resonances -- similar to the well-known gradient threshold for the (slab)
gyrokinetic ITG modes at very small frequencies. Cauchy's argument principle
\cite{hasegawa1968theory} implies that threshold parameters for the
instability fulfill the dispersion relation exactly at real
$\omega$. Conversely, stability is implied, if no such threshold parameters
can be found.

Due to the Gaussian dependence of the imaginary part of $Z$, it can be
neglected in all but one Larmor resonance $p$ for frequencies close to the
gyro resonances, i.e., if $\delta \omega=(\omega-p)\ll1$ and $|k_z|\ll 1$, the
latter of which must be verified a posteriori. The imaginary part of
(\ref{eq:dispersionrelation},\ref{eq:ionresponse}) then turns into
\begin{equation}
[-\omega+k_y(\kappa_n+\kappa_T\partial_T)]
\frac{\Gamma_p(T\ko^2)}{\sqrt T}e^{-\frac{\delta\omega^2}{2\kz^2T}}=0,\label{eq:im}
\end{equation}
equivalent to
\begin{equation}
  -\frac{\omega}{k_y}+\kappa_n+
  \left(\frac{\delta\omega^2}{k_z^2}+\frac{d\ln\Gamma_p(\ko^2)}{d\ln\ko}-1\right)\frac{\kappa_T}{2}
  =0,\label{eq:im1}
\end{equation}
while the real part of the dispersion relation is
\begin{equation}
  \chi_{\text{nr}}+\left[-\omega+{k_y}(\kappa_T\partial_T+\kappa_n)\right]
  \frac{\Gamma_p(T\ko^2)}{|\kz|\sqrt{2T}}Z\left(\frac{\delta\omega}{|\kz|\sqrt{2T}}\right)=0\label{eq:re}
\end{equation}
with the weakly $\omega$-dependent non-resonant charge density response
defined as
\begin{equation}
  \chi_{\text{nr}}=-2+\left[-\omega+{k_y}(\kappa_T\partial_T+\kappa_n)\right]
  \sum_{q\in\mathds{Z},q\ne p}\frac{\Gamma_q(T\ko^2)}{|\kz|\sqrt{2T}}Z\left(\frac{\omega-q}{|\kz|\sqrt{2T}}\right).
\end{equation}
In the limit $\omega=p$, $k_z\rightarrow0$ and small gradients, the constant
term dominates the expression for $\chi_{\text{nr}}$. Direct numerical
inspection for varying $p$ and arbitrary $\ko$ shows that
$|\chi_{\text{nr}}+2|< 1.29|1-k_y\kappa_n/\omega|+0.52|k_y\kappa_T/\omega|$
for $p\ne0$, and $|\chi_{\text{nr}}+2|<|k_y\kappa_T|$ for $p=0$. Hence, for
sufficiently small gradients $\chi_{\text{nr}}<0$, and more specifically
$-2.3\lesssim\chi_{\text{nr}}\lesssim-0.7$.  Solving (\ref{eq:im}) for
$\kappa_n$ and inserting it into (\ref{eq:re}) cancels the plasma dispersion
function,
\begin{equation}
  \chi_{\text{nr}}+\frac12\frac{\delta\omega}{k_z^2} \kappa_T k_y\Gamma_p(\ko^2)=0\label{eq:re2},
\end{equation}
where the identity $Z'(x)=-2(1+xZ(x))$ has been used.

Solving (\ref{eq:re2}) for $k_z^2$ one can replace $k_z$ as a parameter by
$\delta\omega$ provided that $\delta\omega\, k_y \kappa_T\chi_{\text{nr}}<0$.
With this, (\ref{eq:im}) results in
\begin{equation}
  -\frac{p}{k_y}+\kappa_n+\left(\frac{d\ln\Gamma_p(\ko^2)}{d\ln\ko}-1\right)\frac{\kappa_T}2
  -\delta\omega\left(1+\frac{\chi_{\text{nr}}}{\Gamma_p}\right)=0.\label{eq:thresh}
\end{equation}
Fixing the signs of $k_y,\kappa_T$ to be positive, the sign of the term
proportional to $\delta\omega$ is always positive for low gradients, since
$\chi_{\text{nr}} <-\Gamma_p$, because $\Gamma_0\le1$, $\Gamma_p<0.25/p$ for any
$\ko$. Thus, the existence of a threshold requires the rest of
(\ref{eq:thresh}) to be negative, in other words,
\begin{equation}
  \kappa_n<\frac{p}{k_y}+\left(1-\frac{d\ln\Gamma_0(\ko^2)}{d\ln\ko}\right)\frac{\kappa_T}2.\label{eq:nongyrothresh}
\end{equation}

For the gyrokinetic ITG mode $p=0$, and the standard threshold condition
results. For the instability of non-gyrokinetic modes (\ref{eq:nongyrothresh})
requires $p>0$ and no conditions on the gradients themselves besides their
assumed smallness, and the condition that $\kappa_T\ne0$, since otherwise
(\ref{eq:re2}) cannot be solved for $k_z^2$. Then, for an arbitrarily small
$\delta \omega$ a marginally stable threshold mode exists at small $k_y,\ko$,
since
\begin{equation}
   \frac{k_y\chi_{\text{nr}}}{\Gamma_p(\ko^2) }=-O( k_y^{1-2p}) \rightarrow-\infty \mbox{
    for }k_y=\ko\rightarrow0
\end{equation}
and another one at large $\ko$ at fixed $k_y$, since
\begin{equation}
 \frac{k_y\chi_{\text{nr}}}{\Gamma_p(\ko^2) }=-O(\ko)\rightarrow+\infty \mbox{
    for }\ko\rightarrow\infty,k_y=const.
\end{equation}
Since according to (\ref{eq:re2}) in both limits $\kz\rightarrow0$, the initial
assumption of a single resonance contributing to the imaginary part of the ion
response is justified.
On the other hand, if (\ref{eq:nongyrothresh}) does not hold,
(\ref{eq:thresh}) has no valid solutions in $\ko$ and $\delta \omega$, and
hence there are no real solutions for any $\omega$, $k_z$, $k_y$, and $\ko$
for the system of equations (\ref{eq:im},\ref{eq:re}) and therefore according
to Cauchy's argument theorem \cite{hasegawa1968theory} there are no
low-gradient instabilities close to the resonance in the system.

Since the condition $p>0$ only concerns the sign of the phase velocity and not
the gradients themselves, the non-gyrokinetic high frequency IBW instabilities
occur {\em even when the gyrokinetic ITG modes are ruled out by the
  thresholds} -- the IBWs just require a non-zero temperature gradient.

\begin{figure}
    \includegraphics[width=1.05\linewidth]{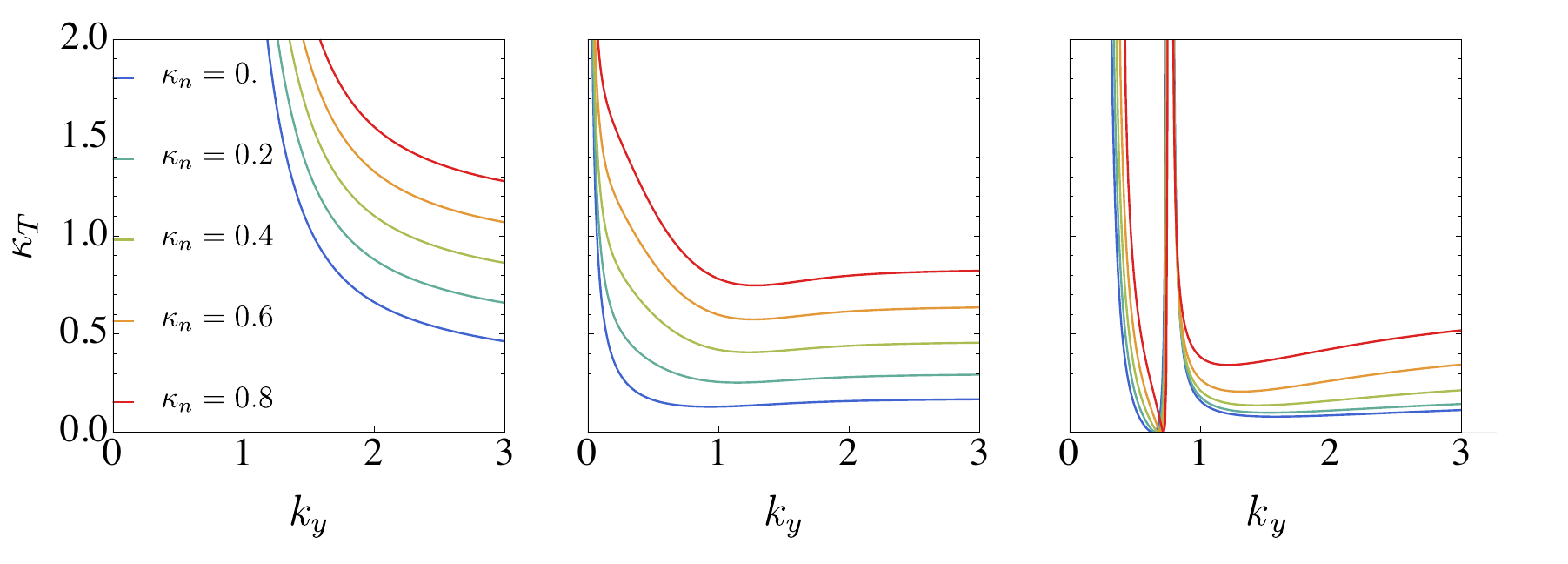}
  	\caption{Critical temperature gradient for low frequency ITG mode
          (middle) and first harmonic IBWs with $p=-1$ (left) and $p=1$
          (right) for varying $\kappa_n$ at $k_z=0.025$.
          \label{fig_stability_criterion}}
\end{figure}

\begin{figure}
  \centering
  \includegraphics[width = 0.4\textwidth]{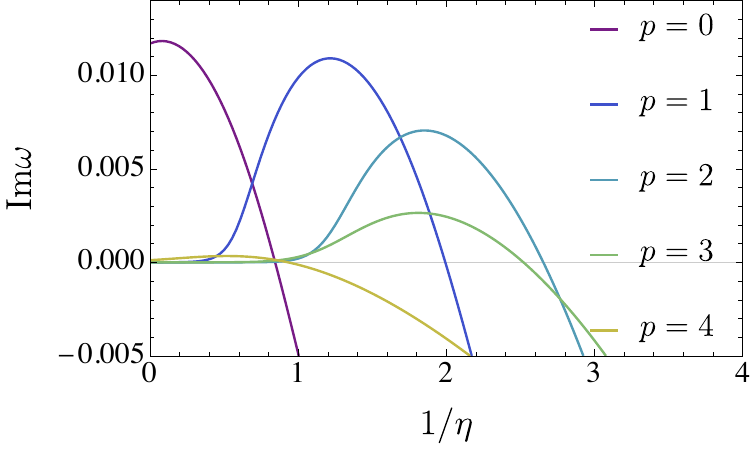}
  \caption{Growth rate of ITG ($p=0$) and IBW ($p\ne0$) instability as a function of
    $\frac 1\eta = \frac{\kappa_n}{\kappa_T}$ for fixed temperature gradient
    $\kappa_T = 0.36$ and $k_y=\ko=2$,
    $k_z=0.025$. \label{fig_growthrate_ratio}}
\end{figure}

Solving the equations (\ref{eq:re2},\ref{eq:thresh}) numerically for
$\delta\omega$ and $\kappa_T$ without restriction to weak gradients
(Fig.~\ref{fig_stability_criterion}) confirms the theory.
%
%
As expected, low frequency slab ITG modes at $p=0$ show an increasing
temperature gradient threshold with increasing density gradient. The ratio of
temperature to density gradient has to be
$\eta = \frac{\kappa_T}{\kappa_n}\gtrapprox 0.9$ at the most unstable
wavenumber. In contrast, the graph for $p=1$, but not $p<0$, shows that the
Bernstein instability can occur for very small temperature gradients.
Interestingly, in the H-mode typically $\eta = 0.6-1.2$
\cite{wolfrum2007edge}, which should largely suppress the gyrokinetic ion
temperature gradient mode.
In all cases the critical temperature gradient increases with the density
gradient.
As for the growth rates at fixed temperature gradient
(Fig.~\ref{fig_growthrate_ratio}) one sees that in contrast to $p=0$ the modes
with $p>0$ are further destabilized with increasing density gradient. The ITG
instability is stabilized for $\eta<1.1$, while the high frequency IBW have
the maximum growth rate at $\eta<1$, again suggesting their relevance in the
H-mode regime, possibly as part of the residual transport.

\paragraph*{6D turbulence simulations}

Having executed extensive linear and nonlinear benchmarking simulations
\cite{mariothesis,raeth_eps} with BSL6D to solve (\ref{eq:vlasov}), confirming
the dispersion relation and reproducing results analogous to
\cite{sturdevant16}, for brevity we report only on a characteristic turbulence
run showing the most relevant features of the high-frequency turbulence.

Since there is no equivalent to the gyrokinetic flux-tube scenario for the
non-gyrokinetic equations without the additional approximation
(\ref{eq:boussinesq}), we use periodic boundary conditions. The ion
distribution function $f(\bx,\bv)$ is initialized with a Maxwellian
$f_0=f_M(n,T)$ with a background density and temperature profile similar to
(\ref{eq:init}),
\begin{equation}
	n = 1 + g_n \sin \frac{2\pi x_{gc}}{L_x}+\text{ran}(\bm r_{gc}),\;
	T = 1 + g_T \sin \frac{2\pi x_{gc}}{L_x},
        \label{eq. temperature_profile}
\end{equation}
to eliminate spurious transient gyro-oscillations, with the addition of a small
white noise density perturbation $\text{ran}(\bm r_{gc})$.
The profile amplitudes have been set to $g_T = 0.6$ and $g_n = 0.49$,
resulting in the maximum inverse gradient lengths $\max(\kappa_T) = 0.375$ and
$\max(\kappa_n) = 0.28$, to specifically render both slab ITG and IBWs
unstable. (For lower temperature gradient the ITG would be completely
suppressed.)

The computation has been carried out in a phase space region covering the
spatial domain $L_x\times L_y\times L_z=4\pi \times 2\pi \times 80\pi$ and the
velocity domain $[-4,4]^3$ using a computational grid of dimensions
$64\times 32\times 16\times 32\times 32 \times 16$ for the phase space
coordinates $(x,y,z,v_x,v_y,v_z)$. $N_t = 1250000$ steps have been carried out
with the time step $\Delta t = 0.008$ up to the end time $t_e=10000$, using
$150000$ core hours on the Intel Xeon 8160 CPUs of the Cineca Marconi cluster.
In the following, band pass filters have been used diagnostically to separate
fluctuation frequencies in the range $|\omega-p|<1/2$ for integer $p$.

\begin{figure}
  \centering
  \includegraphics[width = 0.4 \textwidth]{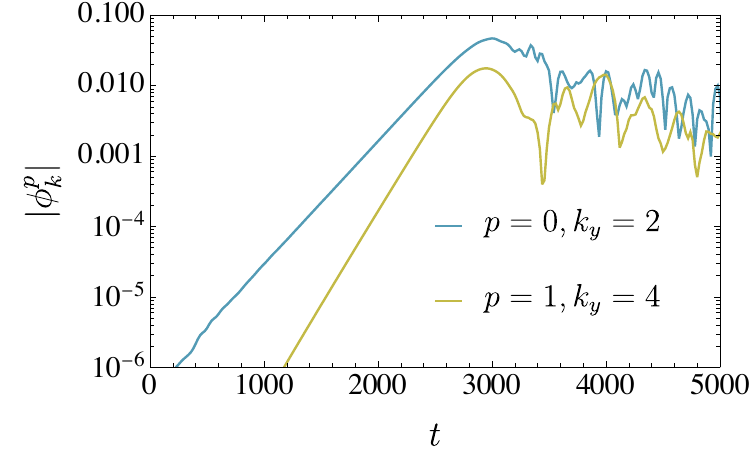}
  \caption{Amplitude of dominant Fourier modes in $y$-direction at their local
    maxima, $x=5.6$ and $x=3.6$ at $k_z=0.025$ for gyrokinetic modes $p=0$ and
    the first Larmor harmonic $p=1$.\label{fig_linear_growth}}
\end{figure}

The simulation starts with the linear growth followed by the turbulent
saturation of the instabilities. The amplitude of the dominant gyrokinetic and
non-gyrokinetic modes are shown in Fig.~\ref{fig_linear_growth}. Their
respective growth rates (indexed ``BSL6D''),
\centerline{
  \begin{tabular}{lcc}
    &\mbox{\quad$\gamma_{\mathrm{BSL6D}}$\quad}&$\gamma_{\mathrm{ana}}$\\
    \hline
    $p=0$&0.0040&0.0024\\		
    $p=1$&0.0062&0.0070
  \end{tabular}}
are consistent with the ones (indexed ``ana'') computed from the dispersion
relation (\ref{eq:dispersionrelation}) using the local gradients
$\kappa_n=0.25$, $\kappa_T=0.34$, and $(k_x,k_y,k_z,p)$ equal to
$(0.5,2,0.025,0)$ and $(3,4,0.025,1)$, respectively. An exact agreement cannot be
expected, since the inherently non-local setup of the simulation prohibits
eigenmodes with a well defined wavenumber $k_x$.

\begin{figure}
  \centering
	\includegraphics[width =0.5 \textwidth]{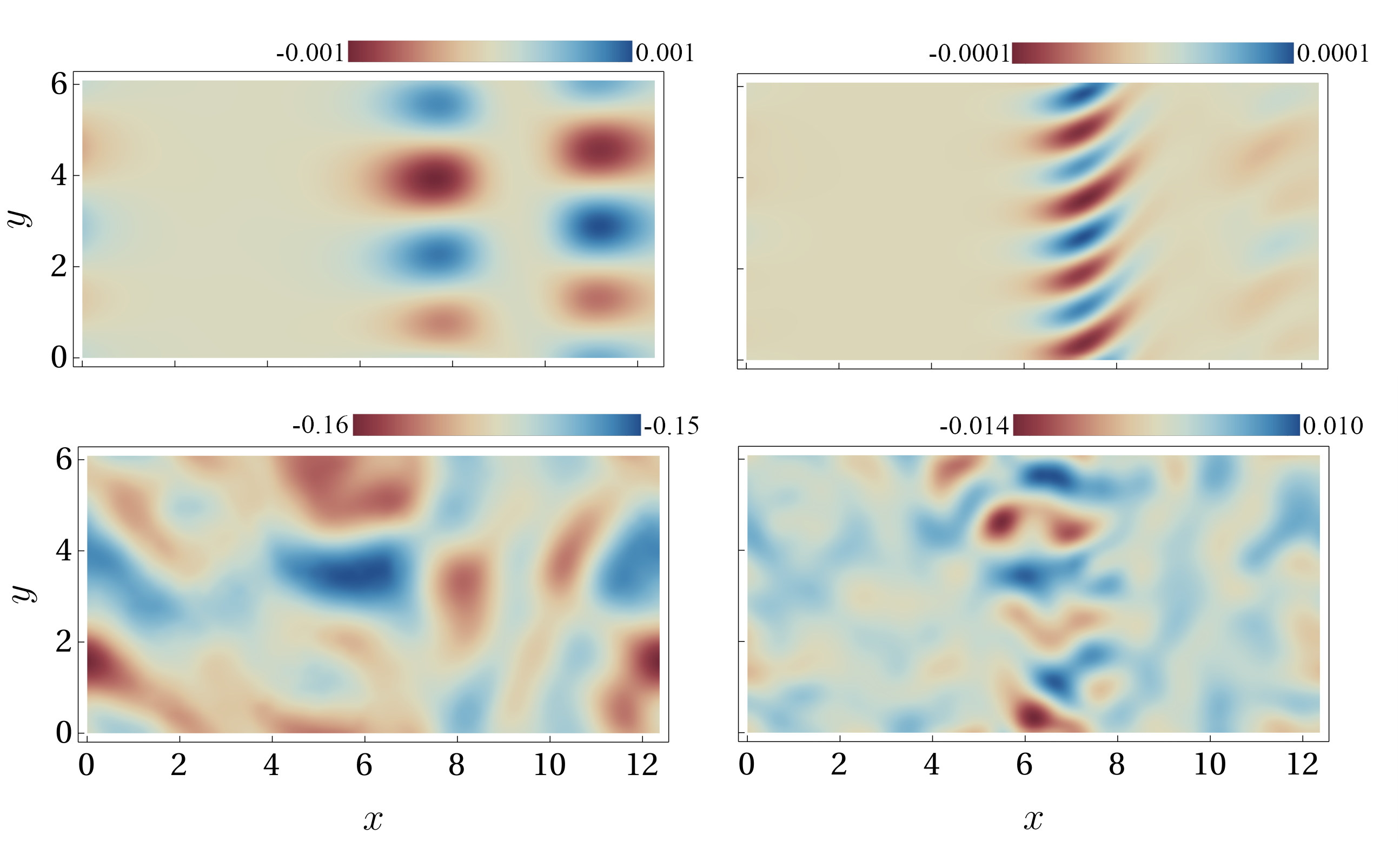}
  \caption{Cross sections of potential fluctuations $\phi$ for $p=0$ (left column) and $p=1$ (right column) at fixed $z$ and $t=1500,5500$ (rows). \label{fig_snapshots_electrostatic_potential} }
\end{figure}

The progression to turbulence is also observable in 2D cross sections of the
electric potential (Fig.~\ref{fig_snapshots_electrostatic_potential}) (for a
movie see \cite{supplementalmovie}). At $t=1500$ the simulation is in the
linear phase and a clear mode structure of the ITG and IBW (with larger
wavenumbers) is visible.  At $t=5500$ the potential indicates nonlinear
ITG/IBW turbulence.

\begin{figure}
  \centering
  \includegraphics[width = 0.40 \textwidth]{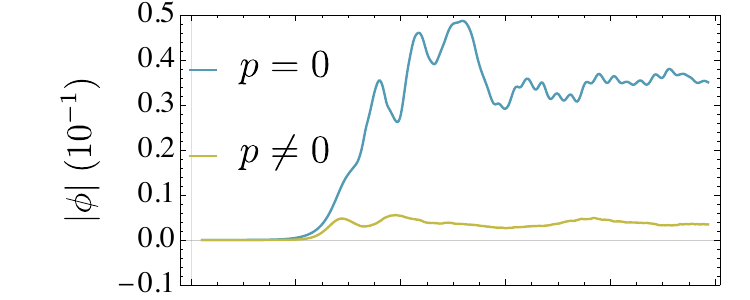}
  \includegraphics[width = 0.4 \textwidth]{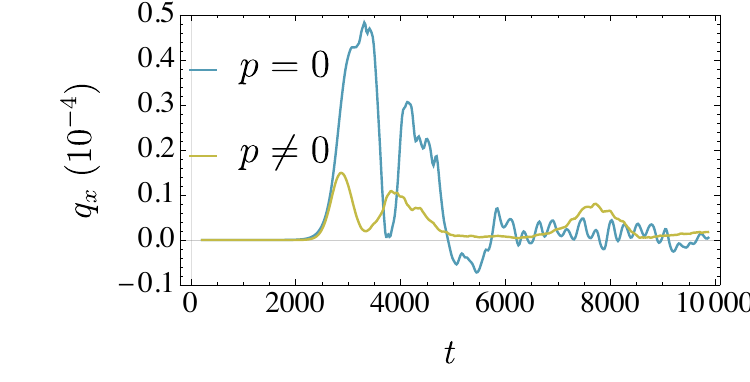}
  
  \caption{Comparison between low frequency modes $p=0$ and IBWs ($p\neq1$)
    for potential amplitude $|\phi|$ (top) and internal ion energy flux
    $q_x$ (bottom), spatially averaged at $x = 7.3$ and box-car averaged
    over $380$ time units. \label{fig_non_lin_saturation}}
\end{figure}

While the total energy flux across the magnetic field in the 6D turbulence
simulations necessarily contains contributions from the Poynting flux
\cite{stringer1991different} and genuinely non-gyrokinetic terms, the dominant
component in the case at hand still comes from the $\bm E\times\bm B$ flow of
the internal ion energy
density $\epsilon=\int fv^2d^3v/2$, namely
$\bm q_\perp\approx\epsilon \hat z \times\nabla\phi.$
Despite their high wavenumbers, the non-gyrokinetic $p\ne0$ IBW components of
the potential fluctuations are significant in comparison to the ITG
fluctuations. The transport of the former even exceeds the one of the latter
(Fig.~\ref{fig_non_lin_saturation}) in the turbulent phase, although the
mixing length argument would suggest a much lower transport due to the high
wavenumber. We believe this to be at least partially due to the (necessarily)
non-local setup, which tends to damp long wavelengths.

\paragraph*{Discussion}
Using a novel 6D kinetic simulation code specifically designed for strongly
magnetized plasmas we have shown that high frequency non-gyrokinetic
instabilities (ion Bernstein waves, IBWs) can produce transport-relevant
turbulence, even in competition to gyrokinetic ITG turbulence, which is
usually exclusively considered in those plasmas.
The simulations have been simplified by assuming electrostatic fields,
adiabatic electrons and a homogeneous magnetic field. They refer therefore
mostly to the ion turbulence component of low-$\beta$, high-gradient plasmas,
where the magnetic fluctuations are small and the magnetic field
inhomogeneities are unimportant compared to the plasma gradients.
Nevertheless, the IBWs can be analytically shown to be unstable at arbitrarily
low gradients and wavenumbers. They require only the presence of a temperature
gradient, and not the excess of a temperature gradient threshold like the ITG
modes.

Gradient lengths such as in the simulation detailed above, $\sim3 \rho_i$, are
regularly encountered in the edge of tokamak H-mode discharges
\cite{wolfrum2007edge}, where also the temperature gradients are typically
below the ITG instability threshold, since
$\eta=L_n/L_T\sim0.6-1.2$.
This, together with the potential breakdown of gyrokinetic theory for short
gradient lengths and the lack of a satisfying explanation for the
L/H-transition, suggests a full 6D treatment of turbulence to be very relevant
to the edge region of magnetic confinement devices.
In future work, more detailed electron physics and inhomogeneous, fluctuating
magnetic fields will be included in such simulations. It seems likely that the
inclusion of additional degrees of freedom would rather increase than decrease
the intensity of the non-gyrokinetic modes.










\begin{acknowledgments}
This work has been carried out partly within the framework of the EUROfusion
Consortium, funded by the European Union via the Euratom Research and Training
Programme (Grant Agreement No 101052200 – EUROfusion). Support has also been
received by the EUROfusion High Performance Computer (Marconi-Fusion). Views
and opinions expressed are however those of the author(s) only and do not
necessarily reflect those of the European Union or the European
Commission. Neither the European Union nor the European Commission can be held
responsible for them.  Numerical simulations were performed at the
MARCONI-Fusion supercomputer at CINECA, Italy and at the HPC system at the Max
Planck Computing and Data Facility (MPCDF), Germany.
\end{acknowledgments}

\end{document}